\UseRawInputEncoding
\documentclass[12pt]{iopart}
\usepackage{amssymb}
\usepackage{floatrow}
\usepackage{graphicx}
\usepackage{cite}
\usepackage[label font=bf]{subfig}         
\usepackage[font=normalsize]{caption}      
\begin{document}
\title{Plasma zone plate for high-power lasers driven by a Laguerre-Gaussian beam}

\author{Lukai Wang$^1$, Wei Liu$^1$, Qing Jia$^{1,*}$ and Jian Zheng$^{1,2,*}$}

\address{$^1$ Department of Plasma Physics and Fusion Engineering and CAS Key Laboratory of Geospace Environment, University of Science and Technology of China, Hefei, Anhui 230026, People’s Republic of China}
\address{$^2$ Collaborative Innovation Center of IFSA, Shanghai Jiao Tong University, Shanghai 200240, People’s Republic of China}
\ead{qjia@ustc.edu.cn and  jzheng@ustc.edu.cn}
\vspace{10pt}
\begin{indented}
\item[]September 2022
\end{indented}

\begin{abstract}
Plasma-based optics has emerged as an attractive alternative to traditional solid-state optics for high-power laser manipulation due to its higher damage threshold. In this work, we propose a plasma zone plate (PZP) driven by the ponderomotive force of a Laguerre-Gaussian beam when it irradiates an underdense plasma slice. We formulate the theory of the PZP and demonstrate its formation and function of focusing high-power lasers using particle-in-cell simulations. The proposed scheme may offer a new method to focus and manipulate high-power lasers with plasma-based optics.
\\ \hspace*{\fill} 
\end{abstract}

%
%
%
%
%

\section{Introduction}
In the last few decades, the rapid development of revolutionary laser techniques, such as chirped pulse amplification \cite{RN1} and optical parametric chirped pulse amplification \cite{RN2}, has significantly improved the peak intensity of lasers and opened up new avenues for fundamental research in fields such as strong field quantum electrodynamics \cite{RN3,RN4}, high-energy-density physics \cite{RN5} and relativistic optics \cite{RN6}. The highest laser intensity available at the moment is delivered by high-power petawatt (PW) lasers \cite{RN7} and can realize intensities up to $\sim{10}^{23}W/cm^2$ after being focused \cite{RN8}. With the increasing peak intensity of lasers, manipulation of such high-power lasers has become increasingly challenging because the size of the traditional solid-state optical components must be enlarged to avoid laser-induced thermal damage \cite{RN9}, which can be extremely costly and technically challenging for large-scale PW laser systems \cite{RN10}. As a result, plasma-based optical components composed of free electrons and ions, which are not limited by optical damage as the traditional solid-state optical components, have become potential alternatives for high-power laser manipulation and have been extensively studied in recent years \cite{RN11,RN12,RN13,RN14,RN15,RN16,RN17,RN18,RN19,RN20,RN21,RN22,RN23}. A series of plasma-based optical components and applications have been proposed, such as plasma gratings or photonic crystals with band structure \cite{RN11,RN12,RN13}, plasma holograms for focusing and mode conversion \cite{RN14}, plasma mirrors for probing strong field quantum electrodynamics \cite{RN15,RN16}, plasma lenses for laser shaping \cite{RN17}, plasma waveplates for polarization manipulation \cite{RN18,RN19,RN20} and plasma-based ellipsoidal mirror \cite{RN21} and compound parabolic concentrator \cite{RN22} with an intensity toleration over ${10}^{12} W/cm^2$ for focusing lasers. Recently, Edwards \etal demonstrated a highly-efficient plasma-based zone plate by two copropagating lasers which tolerates a maximum intensity of ${10}^{17} W/{cm}^2$ \cite{RN23}. The structure of the zone plate reminds us of the intensity patterns of Laguerre-Gaussian lasers.

Laguerre-Gaussian (LG) beams, the eigensolution of the Helmholtz equation in a cylindrical coordinate system under the paraxial condition, are characterized by the azimuthal mode index $l$ and the radial mode index $p$, where $l$ denotes the number of $2\pi$ phase cycles around the circumference and $(p+1)$ denotes the number of radial nodes in the mode profile \cite{RN24}. Almost all previous works on the interaction between LG lasers and plasma focus on optical angular momentum related properties, which are characterized by the azimuthal mode index $l$ \cite{RN25,RN26,RN27}. The intensity pattern of an LG beam with a nonzero $l$ consists of $p$ halos separated from each other by dark rings and a dark center. 

In this work, we propose a plasma zone plate (PZP) driven by the ponderomotive force generated by an LG beam with a nonzero $p$. The formation and functioning of PZP are considered theoretically and verified by numerical and particle-in-cell (PIC) simulations. The merit of using an LG beam is that as an eigenstate in vacuum, the distance of stable propagation is longer and the engineering complexity is lower compared with a non-eigenstate beam, which makes it more plausible for potential experiments. This study extends the understanding of the plasma density evolution driven by the ponderomotive force, which could be of great benefit to both theoretical and experimental works of related research. 

The paper is organized as follows: Proof-of-principle demonstrations of our scheme by 3D-PIC simulations are shown in section 2. The theoretical model of PZP formation is presented in section 3. The formation process is divided into three main stages and a detailed comparison between the theoretical and simulation results is demonstrated. The functioning of the PZP is specified in section 4. On the one hand, the theoretical model and typical examples for focusing infrared and terahertz (THz) lights are presented. On the other hand, the more realistic scenario of an obliquely incident probe beam is discussed. Conclusions are summarized in section 5.

\section{Proof-of-principle PIC demonstrations}

First of all, we present proof-of-principle demonstrations of our scheme using 3D-PIC simulations by the code EPOCH \cite{RN28} to illustrate the formation and focusing function of a typical PZP. The schematic is shown in figure 1(a).

A circularly polarized (CP) LG beam ($\lambda_p=1\mu m, l=2, p=4$), the intensity of which is shown in figure 1(b), is launched from the –z boundary as the pump beam to irradiate a thin plasma slab for demonstration. The beam radius is $w_p=10\mu m$ and the maximum intensity is $2.7 \times {10}^{16} W/{cm}^2$. The plasma slab consists of a constant density slab of length $2\lambda_p$ with Gaussian density ramps of length $1\lambda_p$ at each side along the z-direction while uniformly distributed on the x-y plane. The maximum density is set to be $n_{e0}=0.3n_c(\lambda_p)$, where $n_c(\lambda_p)[{cm}^{-3}]=1.12\times{10}^{21}/\lambda_p^2[\mu m]$ is the critical density for the laser with wavelength $\lambda_p$. We assume a fully-ionized electron-ion plasma with the ions set to be $q_i=e$ and $m_i=100m_e$ ($e$: the elementary charge, $m_e$: the electron mass) to reduce the simulation time. The plasma temperature is $T_e=25eV$, $T_i=2.5eV$. The dimensions of the simulation box are $L_x \times L_y \times L_z = 160 \mu m \times 160 \mu m \times 40 \mu m$ and the spatial resolutions are $\Delta x = \Delta y = 0.25 \mu m$ and $\Delta z = 0.1 \mu m$. The plasma slab is placed at the center of the simulation box and padded on both sides with sufficiently long vacuum regions. 

\begin{figure}
\sidesubfloat[]{\label{(a)}\includegraphics[width=0.925\linewidth]{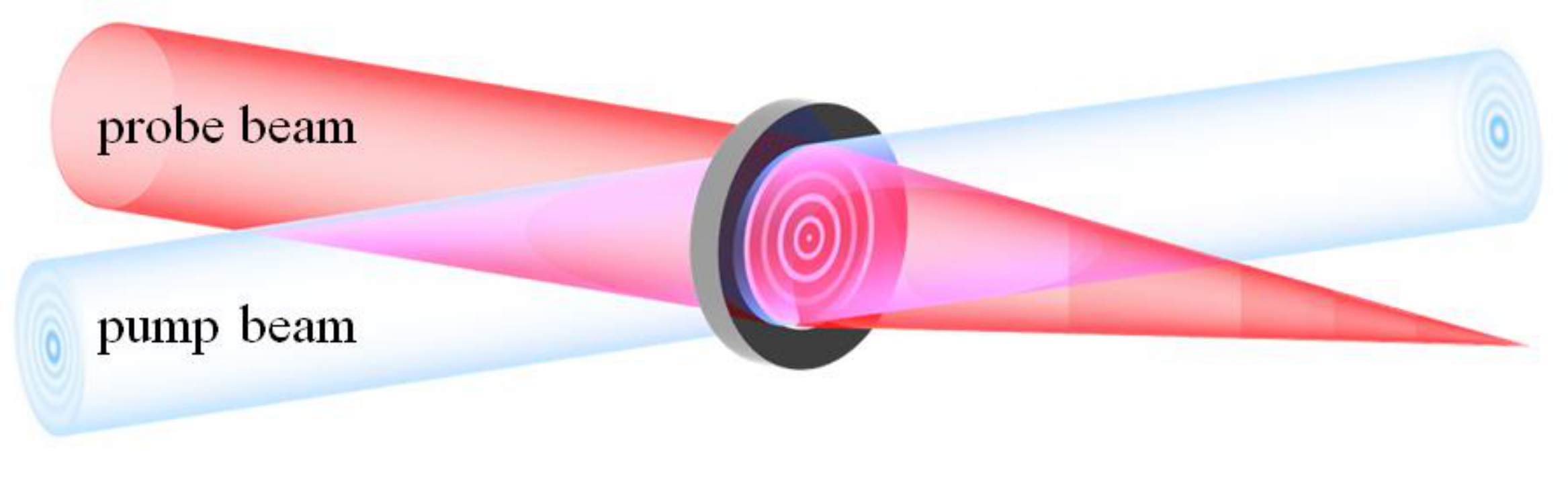}}
\vfil
\sidesubfloat[]{\label{(b)}\includegraphics[width=0.25\linewidth]{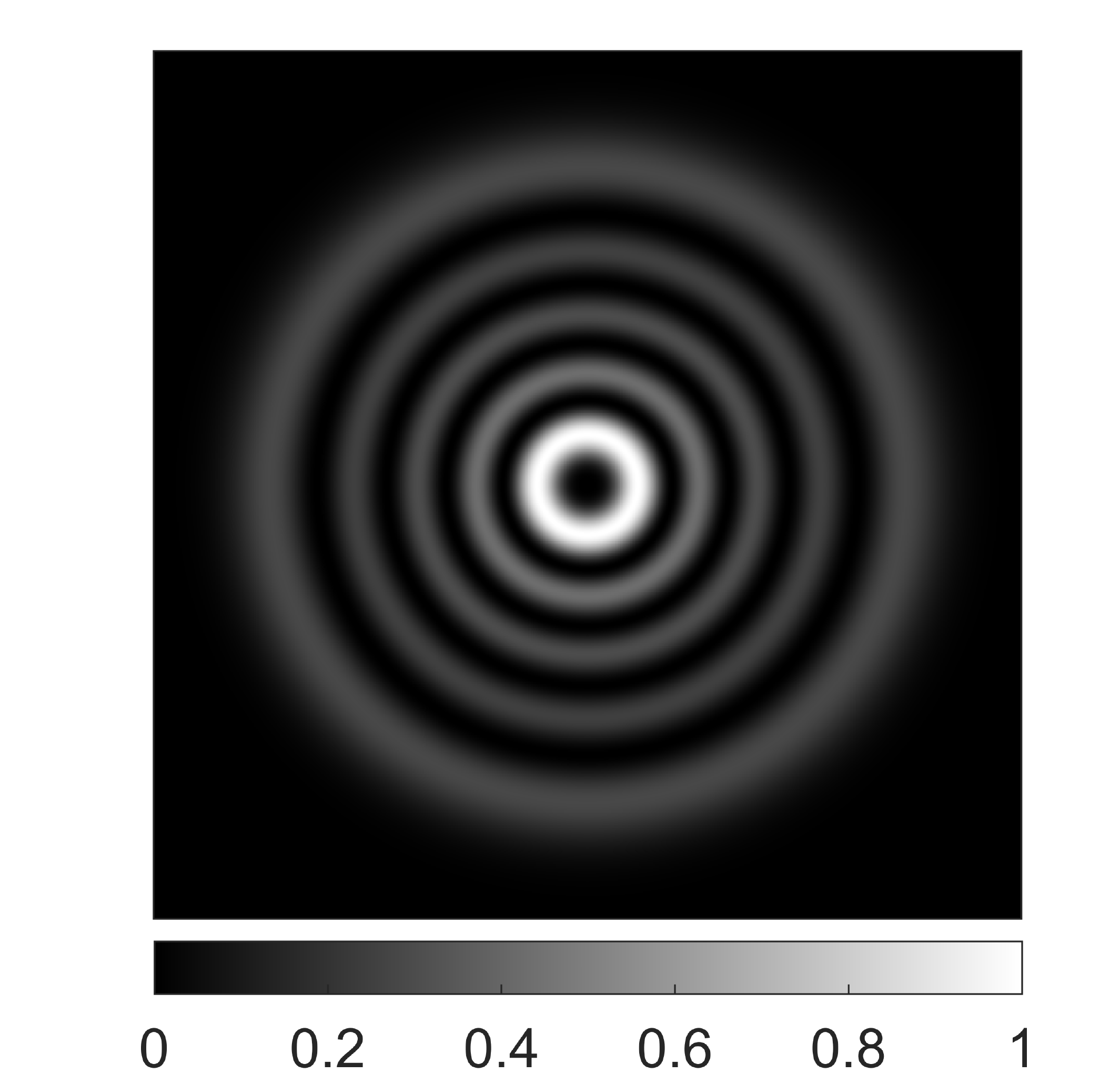}}
\hfil
\sidesubfloat[]{\label{(c)}\includegraphics[width=0.25\linewidth]{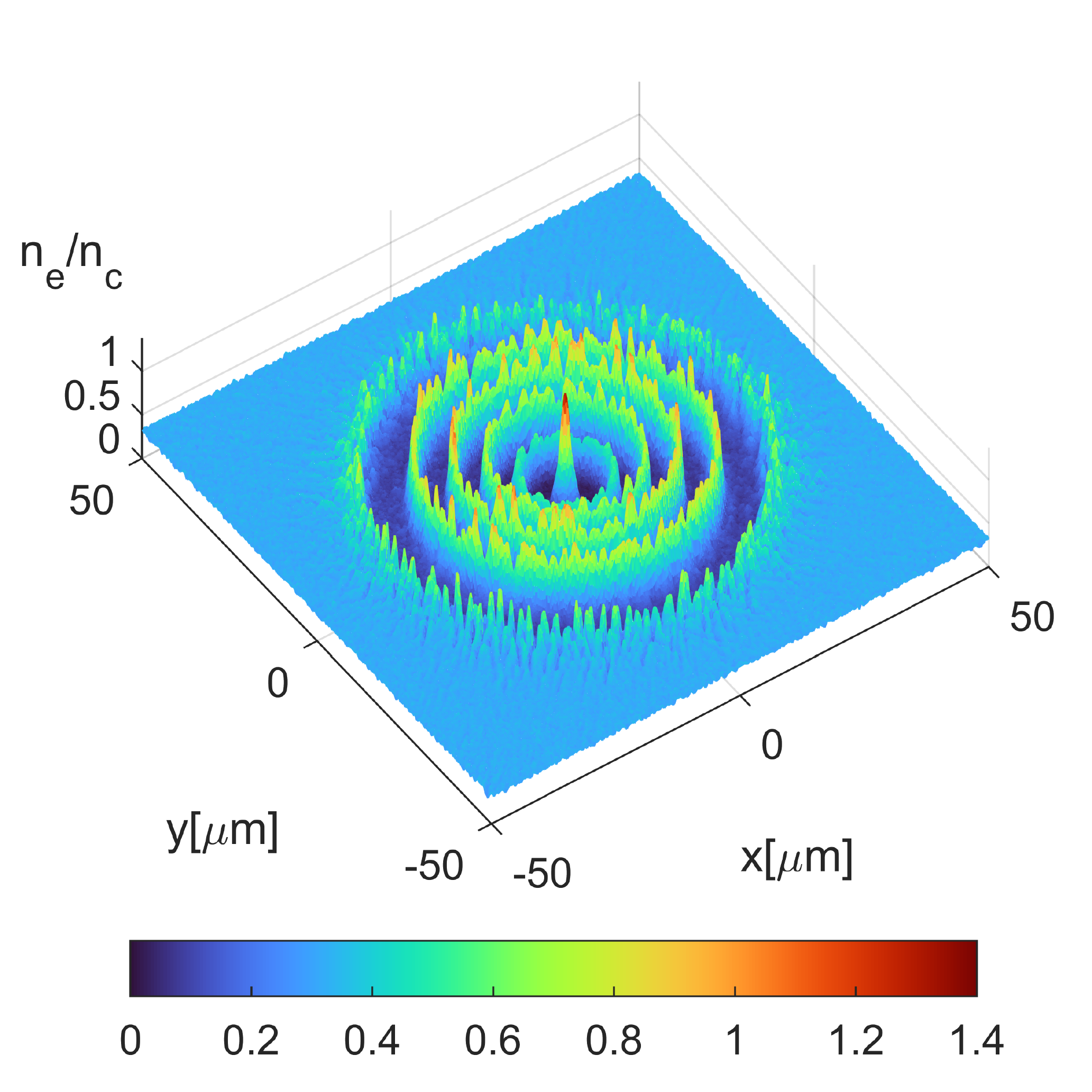}}
\hfil
\sidesubfloat[]{\label{(d)}\includegraphics[width=0.3\linewidth]{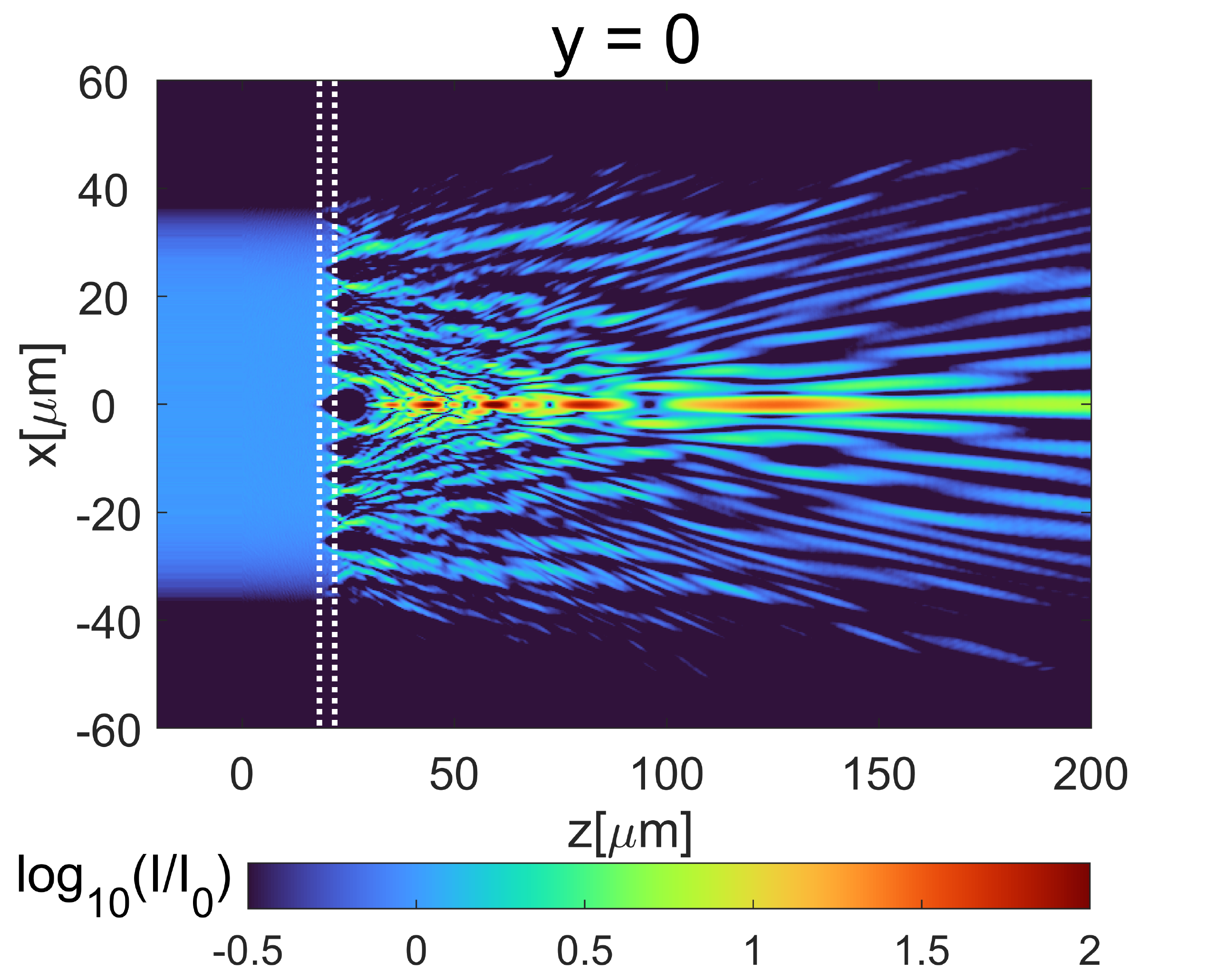}}
\caption{(a) Schematic of a PZP. An LG beam ($l=2,p=4$) is launched into the plasma and generates a PZP. At a delayed time, a probe beam is focused by the PZP. (b) Intensity pattern for the pump beam. (c) Transverse density distribution of the PZP. (d) The laser intensity of the probe beam before, during, and after the interaction with PZP. The white dotted lines ($z=18.2/21.8\mu m$) represent the position of the PZP.}
\end{figure}

As the pump beam interacts with the plasma, an annular density modulation starts to develop and reaches the maximum at around $4ps$, which is shown in figure 1(c). Then a CP probe beam ($\lambda_b=1\mu m$) is launched into this modulated plasma slice. Figure 1(d) shows the intensity distribution of the probe beam through its propagation. The modulated plasma functions as a plasma zone plate (PZP), which magnifies the intensity of the incoming probe beam up to approximately two orders of magnitude at several foci, as can be seen in figure 1(d). Such a multi-focal laser field structure is the typical intensity structure after traversing the Fresnel zone plate (FZP). In the following, we formulated the theoretical model for the formation of PZP and verified it by simulations in detail.

\section{Formation of PZP}
Considering a non-relativistic CP LG laser traveling in the z-direction as the pump beam through an initially homogeneous underdense plasma consisting of electrons (density $n_{e0}=0.3n_c$) and ions (charge $Ze$ and density $n_{i0}=n_{e0}/Z$), in the cylindrical coordinate system, the complex amplitude of the electric field of the laser can be expressed as

\begin{figure}
\sidesubfloat[]{\includegraphics[width=0.45\linewidth]{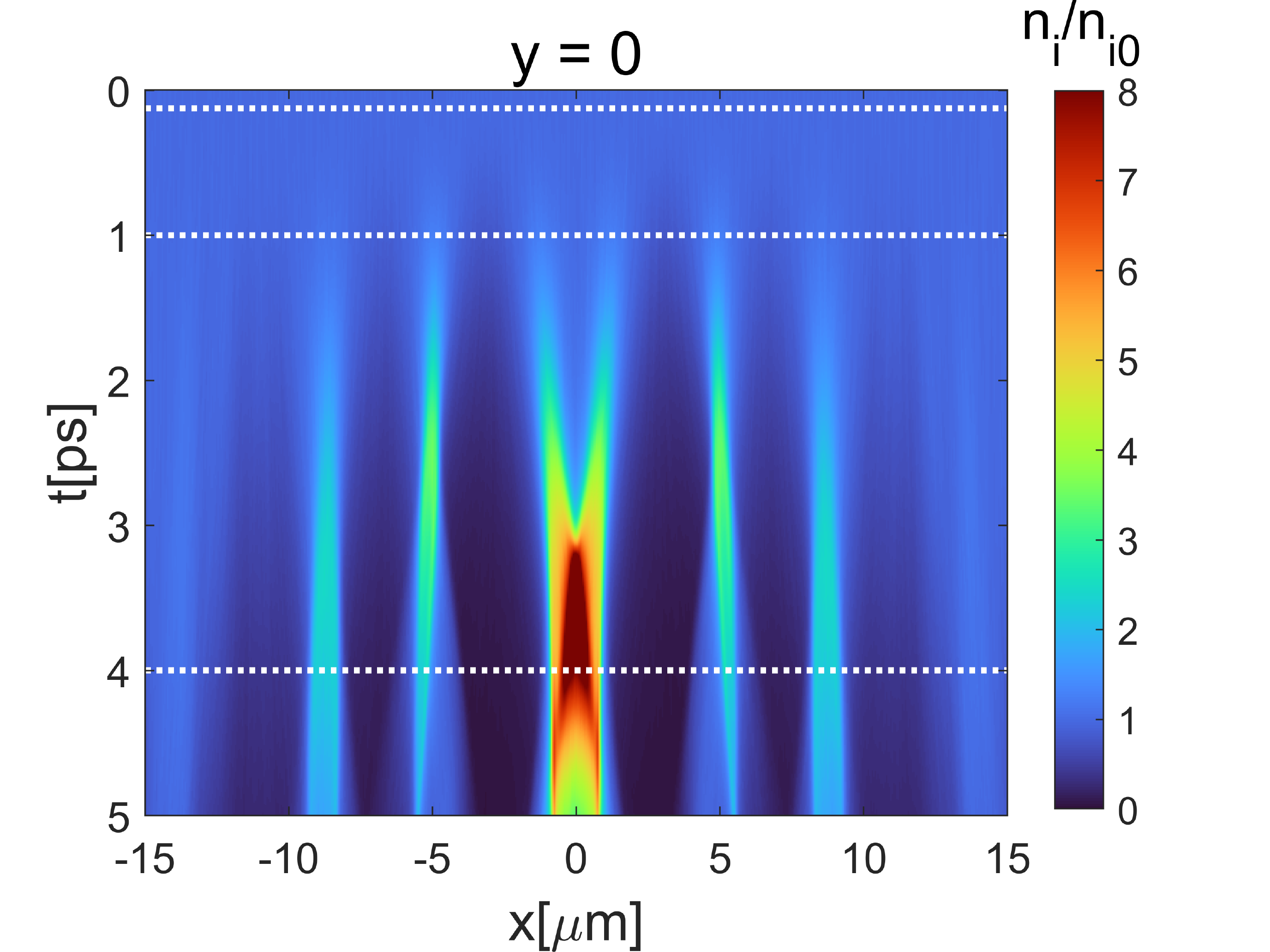}}
\hfil
\sidesubfloat[]{\includegraphics[width=0.45\linewidth]{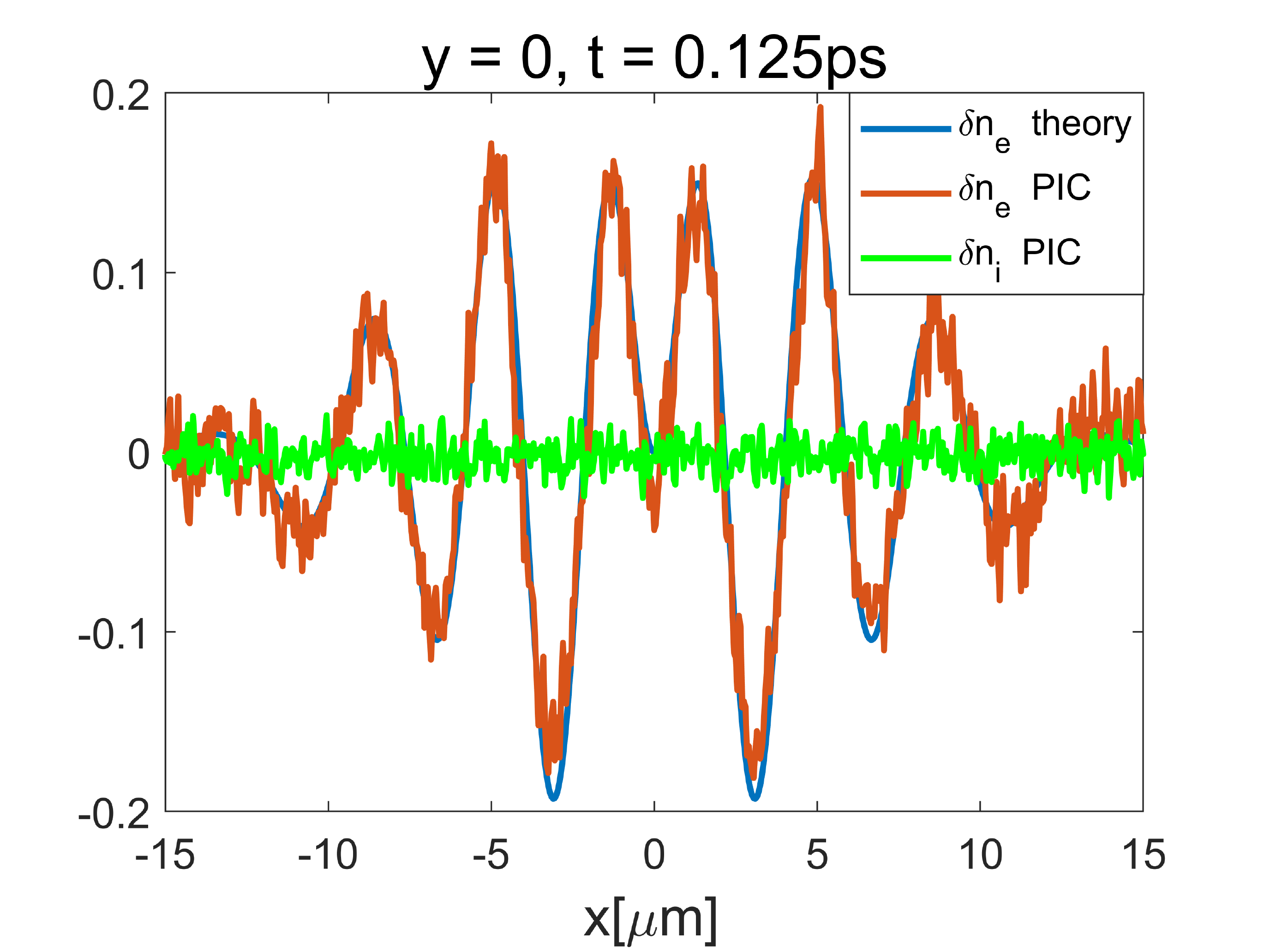}}
\vfil
\sidesubfloat[]{\includegraphics[width=0.45\linewidth]{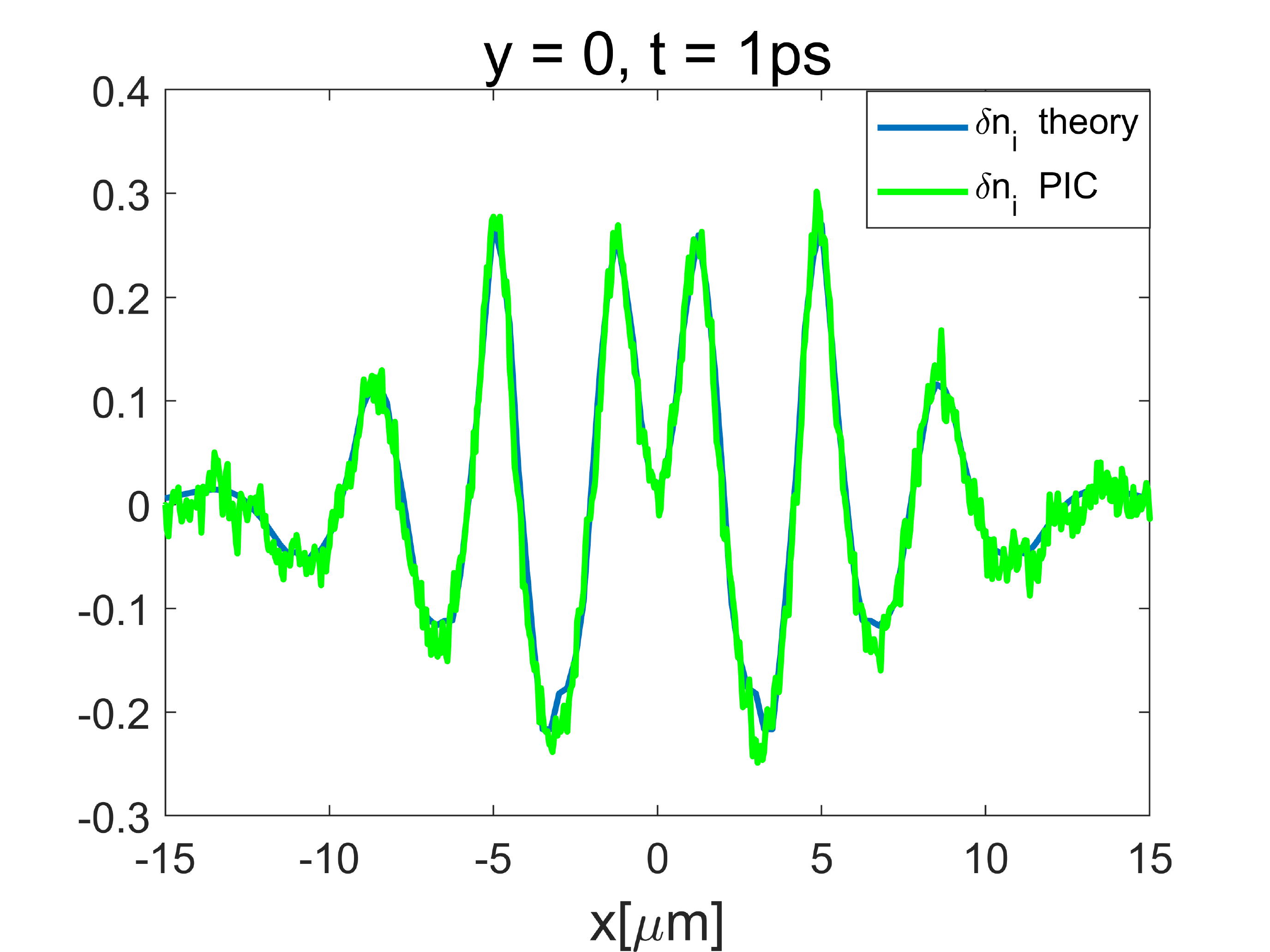}}
\hfil
\sidesubfloat[]{\includegraphics[width=0.45\linewidth]{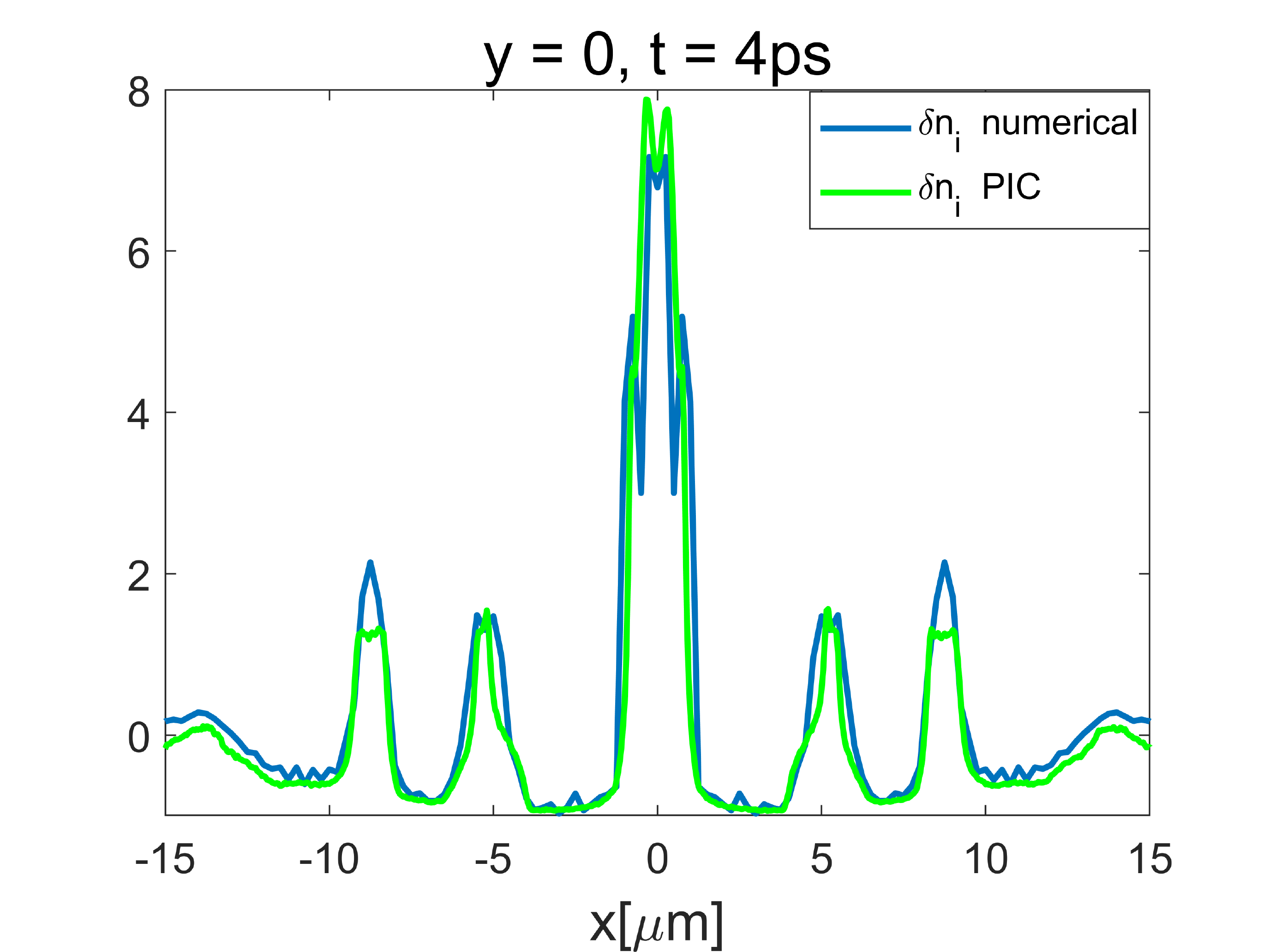}}
\caption{Example PIC simulation: pump beam: $\lambda_p = 1 \mu m,l=2,p=2,w_p= 5 \mu m$, plasma: $n_{e0} = {10}^{-4}n_c(\lambda_p)$. The initial density in the example simulation is set to be as low as $n_{e0} = {10}^{-4}n_c(\lambda_p)$ to obtain an observable electron density variation in stage \uppercase\expandafter{\romannumeral1} according to equation (\ref{eq6}). (a) Transverse evolution of the ion density. The white dotted lines indicate the characteristic times of stage \uppercase\expandafter{\romannumeral1} ($t=0.125ps$), stage \uppercase\expandafter{\romannumeral2} ($t=1ps$) and stage \uppercase\expandafter{\romannumeral3} ($t=4ps$). (b) Blue line: theoretical electron density variation of stage \uppercase\expandafter{\romannumeral1}. Orange line: electron density variation from PIC simulation. Green line: ion density variation from PIC simulation. (c) Blue line: theoretical ion density variation of stage \uppercase\expandafter{\romannumeral2}. Green line: ion density variation from PIC simulation. (d) Blue line: ion density variation from numerical simulation. Green line: ion density variation from PIC simulation.}
\end{figure}

\begin{eqnarray}
E_{x,y}=\frac{E_p}{C_p^l}exp(-\frac{r^2}{w_p^2})(\frac{\sqrt{2}r}{w_p})^lL_p^l(\frac{2r^2}{w_p^2})exp[i(kz+l\phi + \varphi_{x,y})] \label{eq1}
\end{eqnarray}
where $E_p$ is the peak amplitude of the electric field, $C_p^l$ is the normalization constant, $L_p^l$ is the Laguerre polynomial, $\varphi_{x,y}$ is the initial phase of $x/y$ polarization, and $k$ is the wavenumber. Here, our analysis is within the Rayleigh range so that the diffraction can be neglected \cite{RN29}.The LG laser intensity is only related to $r$, i.e., $I=I_0f(r)$ where $I_0=\sqrt{\epsilon/\mu}E_p^2$ is the maximum intensity ($\epsilon$ is the permittivity and $\mu$ is the magnetic permeability) and $f(r)$ is the normalized distribution function. Introducing the dimensionless laser amplitude $a_p=eE_p/\omega m_ec$, where $\omega$ is the laser frequency and $c$ is the speed of light in vacuum, the ponderomotive force on the electrons can be expressed by the negative gradient of the ponderomotive potential $\phi_p=\frac{a_p^2}{2}m_ec^2f(r)$ \cite{RN30}:
\begin{eqnarray}
\mathbf{F_p^e}=-\nabla\phi_p  
\label{eq2}
\end{eqnarray}

The ponderomotive force on the ions is negligible since it is smaller than that on the electrons by a factor $Z^2m_e/m_i$ \cite{RN31}. Let $\tilde{n}_\alpha = n_\alpha - n_{\alpha0}$ and $\delta{n_\alpha} = \tilde{n}_\alpha / n_{\alpha0}$ ($\alpha=i,e$) be the absolute and normalized density variation. The temporal evolution of the density can be divided into three stages. Stage \uppercase\expandafter{\romannumeral1} is the beginning of the laser-plasma interaction: ions are assumed to be stationary and the normalized electron density is much less than unity, i.e. $\delta n_i\approx0, \delta n_e\ll1$. During this stage, an electrostatic charge-separation field is generated to balance the ponderomotive force. In stages \uppercase\expandafter{\romannumeral2} and \uppercase\expandafter{\romannumeral3}, the ions move together with the electrons driven by the electrostatic charge-separation field. We define the period when $\delta n_i,\delta n_e\ll1$ as stage \uppercase\expandafter{\romannumeral2} and the period when $\delta n_i,\delta n_e\gtrsim1$ as stage \uppercase\expandafter{\romannumeral3}. In the following, we formulate the theoretical model for the density evolutions in each stage, and verify it with an example simulation, the results of which are shown in figure 2. 

In stage \uppercase\expandafter{\romannumeral1}, the ions are assumed to be stationary due to their large inertial mass. The thermal pressure on electrons $\nabla p_e = \nabla(n_eT_e)$ is negligible because $\delta n_e\ll1$ and the convective term of electrons is negligible as well. Therefore, the fluid equations for electrons can be linearized as:
\begin{eqnarray}
\frac{\partial\tilde{n}_e}{\partial t}+n_{e0}\nabla\cdot\mathbf{v_e}=0  
\label{eq3}
\end{eqnarray}
\begin{eqnarray}
\frac{\partial\mathbf{v_e}}{\partial t}=\frac{e}{m_e}\nabla\phi+\frac{\mathbf{F_p^e}}{m_e}  
\label{eq4}
\end{eqnarray}
where $\phi$ is the electrostatic charge-separation field generated by the Poisson equation:
\begin{eqnarray}
\nabla^2\phi = \frac{e}{\epsilon_0}(n_e-Zn_i)   
\label{eq5}
\end{eqnarray}

Combining equations (\ref{eq3}), (\ref{eq4}) and (\ref{eq5}), one finds the normalized electron density variation as:
\begin{eqnarray}
\delta n_e = \frac{a_p^2c^2}{2\omega_{pe}^2}\nabla^2f(r)  
\label{eq6}
\end{eqnarray}
where $\omega_{pe}$ is the electron plasma frequency. The comparison between the theory and PIC simulation at the initial time of $t=0.125ps$ is shown in figure 2(b), where the ions are still almost homogeneous, while the electron density variation in accordance with the theory has been established. 

As the electrostatic charge-separation field is created in stage \uppercase\expandafter{\romannumeral1}, the ions are driven to move together with the electrons, and a cumulative density modulation will be formed in stage  \uppercase\expandafter{\romannumeral2}. During this stage, the electrons are assumed to remain in force equilibrium while the ion density variation becomes non-negligible, thus $\delta n_e(r)-Z\delta n_i(r)=\frac{a_p^2c^2}{2\omega_{pe}^2}\nabla^2f(r)$. From the linearized fluid equations of ions
\begin{eqnarray}
\frac{\partial\tilde{n}_i}{\partial t}+n_{i0}\nabla\cdot\mathbf{v_i}=0  
\label{eq7}
\end{eqnarray}
\begin{eqnarray}
\frac{\partial\mathbf{v_i}}{\partial t}=-\frac{Ze}{m_i}\nabla\phi  
\label{eq8}
\end{eqnarray}
one finds the spatiotemporal evolution of the ion density:
\begin{eqnarray}
\delta n_i = \frac{Za_p^2c^2}{4}\frac{m_e}{m_i}\nabla^2f(r)t^2  
\label{eq9}
\end{eqnarray}

Comparing equations (\ref{eq6}) and (\ref{eq9}), it is recognized that the electron and ion density variations follow the same spatial pattern, which is directly proportional to $\nabla^2f(r)$. Therefore, an annular density variation that grows quadratically with time is obtained, as can be seen from figures 2(b) and 2(c). A comparison between the simulation and the theory of the ion density variation in stage \uppercase\expandafter{\romannumeral2} ($t=1ps$) is presented in figure 2(c) and exhibits a fair conformity. 

As the normalized density variations continue to grow close to unity, the divergence of the particle flux $\nabla\cdot(n_\alpha\mathbf{v_\alpha})$ ($\alpha=i,e$) cannot be further simplified as $n_{\alpha0} \nabla\cdot \mathbf{v_\alpha}$ any more and the thermal pressure becomes non-negligible. Therefore, the linearized approximation is no longer valid and thus the density variations are no longer theoretically analytic, we define this stage as stage \uppercase\expandafter{\romannumeral3}.

To analyze the density evolutions in stage \uppercase\expandafter{\romannumeral3}, we start from the momentum equations of electrons and ions:
\begin{eqnarray}
n_em_e\frac{d\mathbf{v_e}}{dt} = n_ee\nabla\phi+n_e\mathbf{F_p^e}-\nabla p_e  
\label{eq10}
\end{eqnarray}
\begin{eqnarray}
n_im_i\frac{d\mathbf{v_i}}{dt} = -n_iZe\nabla\phi-\nabla p_i  
\label{eq11}
\end{eqnarray}

Thermal pressure terms are introduced according to the adiabatic law, which implies $p_\alpha n_\alpha^{-\gamma}=const$ with $p_\alpha=n_\alpha T_\alpha$, where $T_\alpha$ is the particle temperature and $\gamma$ is the adiabatic index \cite{RN32}. Combining equations (\ref{eq10}) and (\ref{eq11}) and neglecting electron momentum \cite{RN33} yields
\begin{eqnarray}
m_i\frac{d\mathbf{v_i}}{dt} = -Z\nabla \phi_p - \frac{\nabla(p_e+p_i)}{n_i}  
\label{eq12}
\end{eqnarray}

Thermal pressure does not play an essential role at first while the density variation is small. As the density variation grows continuously, the thermal pressure grows along and reaches the same order of magnitude as the electrostatic force at $|\delta n_i|\sim|\frac{Zm_ec^2a_p^2}{ZT_{e0}+T_{i0}}|$. According to equation (\ref{eq12}), ions will eventually be trapped in the troughs of the ponderomotive potential and so will the electrons. Since the ponderomotive potential is proportional to the laser intensity, we conclude that eventually a density modulation where the particles are trapped at the low intensity nodes of the 
pump beam will be formed. This implies that the density variations in stage \uppercase\expandafter{\romannumeral3} are negatively correlated with $f(r)$, while proportional to $\nabla^2f(r)$ in stage \uppercase\expandafter{\romannumeral1} and \uppercase\expandafter{\romannumeral2}. The transition from stage \uppercase\expandafter{\romannumeral2} to stage \uppercase\expandafter{\romannumeral3} can be found as the density variation develops from proportional to $\nabla^2f(r)$ in figure 2(c) to negatively correlated with $f(r)$ in figure 2(d), which is marked by the emergence of the density peak at $x=0,y=0$ at around $2.5ps$ as shown in figure 2(a). As the density variation in stage \uppercase\expandafter{\romannumeral3} is analytically difficult to resolve, we designed a particle-fluid program \cite{2020Growth} to numerically calculate the ion density evolution based on equation (\ref{eq12}). A comparison of the ion density in stage \uppercase\expandafter{\romannumeral3} ($t=4ps$) between our particle-fluid program and the PIC simulation is presented in figure 2(d). 

The final modulated plasma slice is composed of alternating high and low density rings corresponding to the alternating dark and bright rings of the incident LG beam. According to the dispersion relation of electromagnetic waves in plasma \cite{RN34}, such a density distribution can be interpreted as an annular refractive index structure with alternating low and high refractive index rings, thus we regard the obtained modulated plasma as a PZP. 

\section{Functioning of PZP}
In this section, the focusing effect of PZP on the probe beam is considered based on the laser propagation equation and verified by PIC simulations. Considering a probe beam normally incident on the PZP and approximate the plasma as a typical medium, the refraction of the probe beam is governed by \cite{RN35}:

\begin{eqnarray}
(\frac{\partial^2}{\partial t^2}-c^2\nabla^2)a_b=-\frac{n_ee^2a_b}{\gamma\epsilon_0m_e}  
\label{eq13}
\end{eqnarray}
where $\epsilon_0$ is the vacuum permittivity and $\gamma$ is the relativistic factor given by $\gamma=\sqrt{1+\frac{a_b^2}{2}}$. The corresponding refractive index is given by $n=\sqrt{1-\frac{n_e}{\gamma n_c}}$. Thus, the alternating high and low density rings are alternating low and high refractive index rings for the probe beam. During propagation through the PZP, most of the energy of the probe beam will be refracted to the high refractive index 
rings and a phase shift is generated between adjacent rings. Let $n_l$ and $n_h$ be the refractive indexes of two adjacent rings, $\Delta r$ be the distance between the two rings, $D$ be the thickness of the PZP and $k_b$ be the wavenumber of the probe beam. The characteristic length to produce a significant amplitude modulation can be given by $L=(\frac{n_h^2}{n_l^2}-1)^{-\frac{1}{2}}\Delta r$ \cite{1969self} and the phase shift can be given by $\Delta\phi=(n_h-n_l)k_bD$ \cite{RN23}. Therefore, the probe beam will be modulated into an emergent beam composed of alternating bright and dark rings (when $D \gtrsim L$) together with phase shift, which is similar to that of an FZP, and focused at multiple foci during the further propagation. 

In the PIC verification simulation, a CP probe beam ($\lambda_b=1\mu m$, $a_b=0.5$) with a maximum intensity of $I_0=6.8\times{10}^{17}W/{cm}^2$ is launched into the PZP obtained in section 2 as shown in figure 1(d). The intensity distribution on the optical axis is presented by the blue solid line in figure 3(a), which indicates that the PZP magnifies the intensity of the incoming probe beam up to a maximum of approximately two orders of magnitude at several foci. The maximum intensity obtained is about $9\times{10}^{19}W/{cm}^2$. The spot size in both the transverse and longitudinal directions is on the same order of the probe wavelength, and increases with the focal length in the same pattern as the traditional FZP \cite{RN36}. We conducted a contrast simulation in which the same probe beam is launched on the PZP generated by the pump beam with $p=2$ and obtained similar laser field structures, as shown by the black dashed line in figure 3(a). The result indicates that the intensity amplification increases with the increasing $p$, because LG beams of higher $p$ generate more alternating rings corresponding to zone plates of higher orders. The LG-light-modulated plasma slab functions as an FZP to focus the probe beam.
\begin{figure}
\sidesubfloat[]{\includegraphics[width=0.45\linewidth,height=5cm]{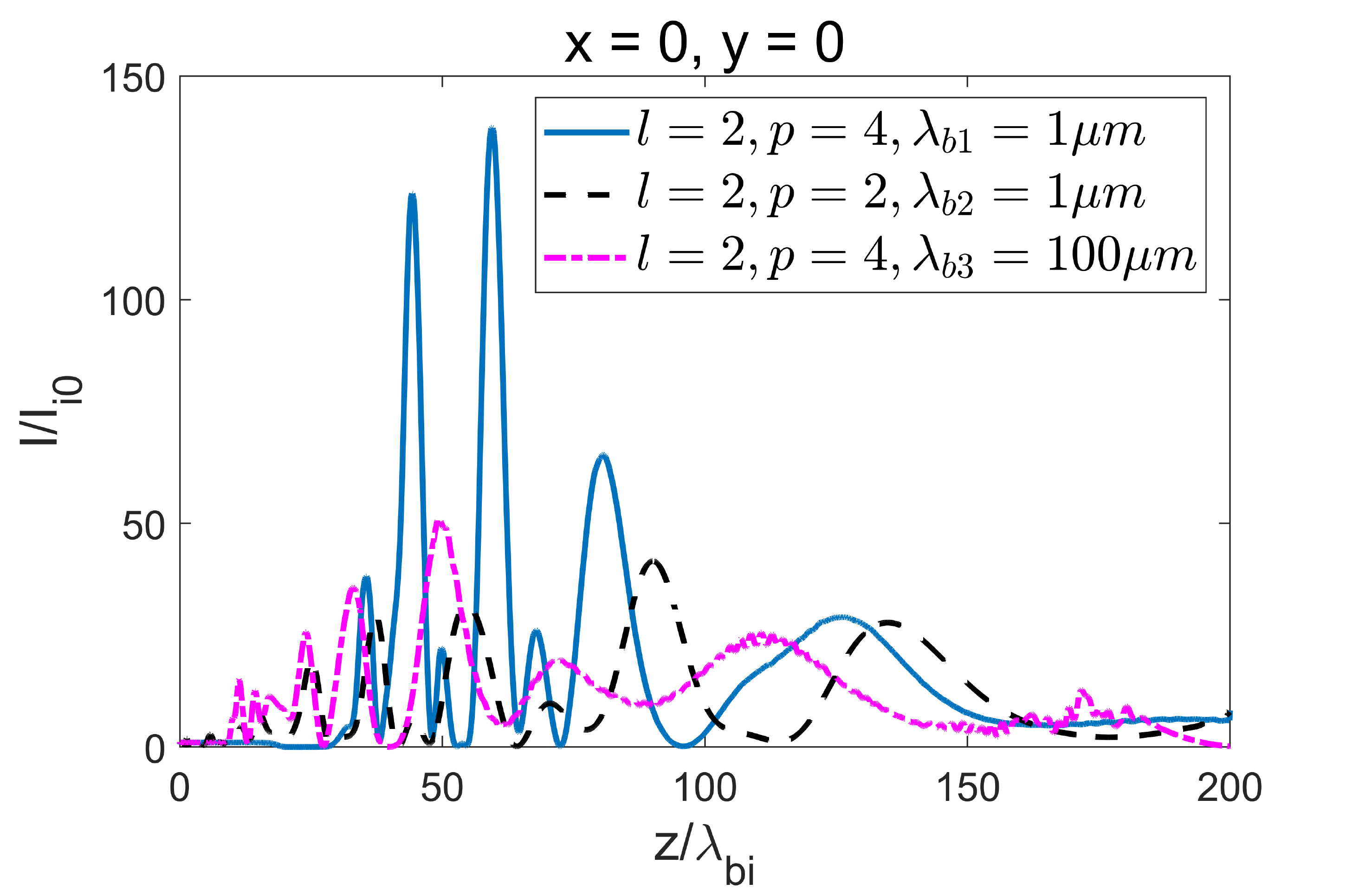}}
\hfil
\sidesubfloat[]{\includegraphics[width=0.45\linewidth,height=5cm]{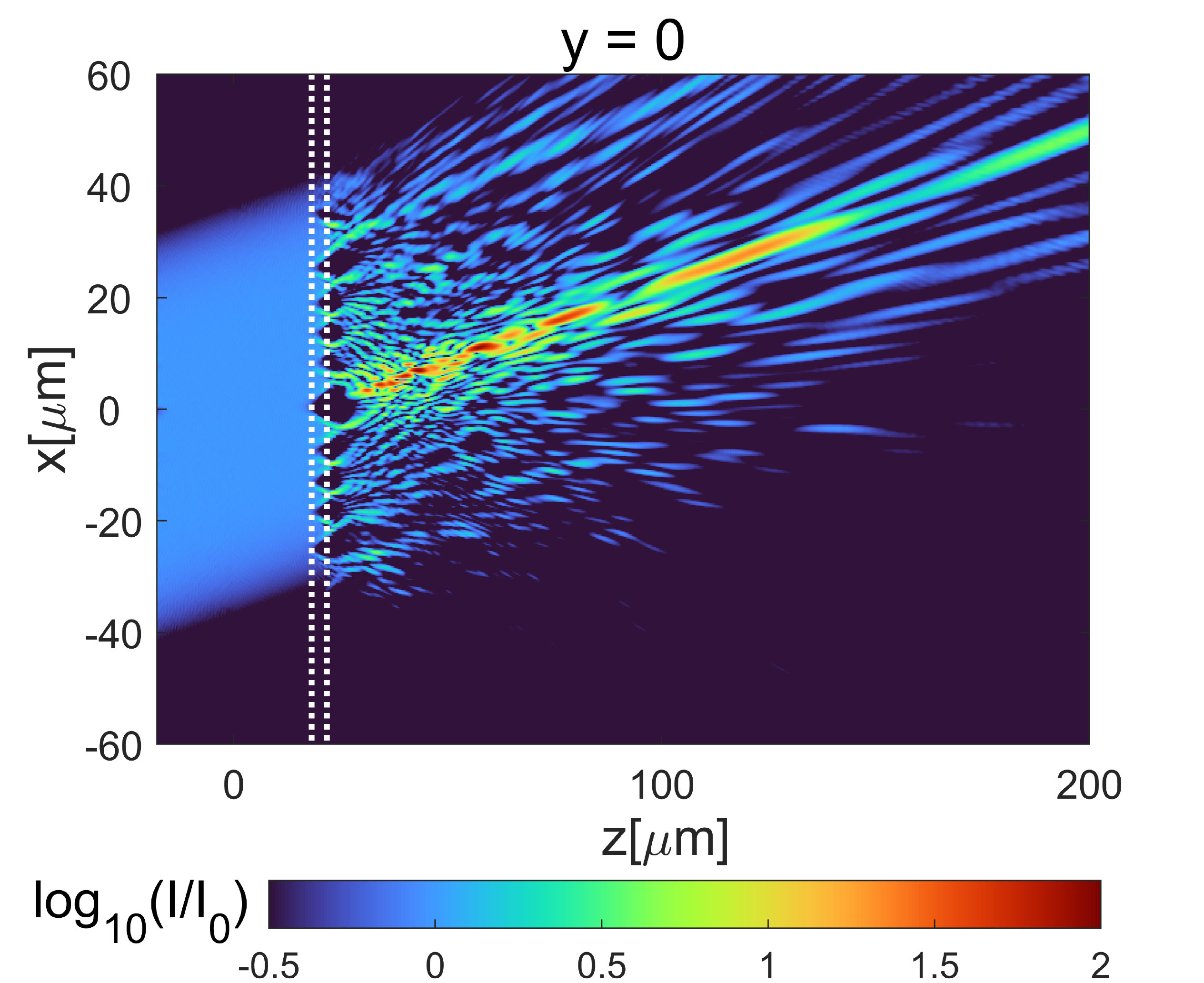}}
\caption{(a) Magnification of the probe beam on the optical axis. Blue solid line: the PZP generated by an $l=2,p=4$. LG laser in section 2. Black dashed line: the PZP generated by an $l=2,p=2$ LG laser. Purple dash-dotted line: THz PZP. (b) The laser intensity of a $15^\circ$ obliquely incident probe beam before, during, and after the interaction with PZP. The white dotted lines represent the location of the PZP.}
\end{figure}

To focus a probe beam with a given wavelength, the plasma density should be on the order of $n_{e}\sim0.1n_c(\lambda_b)$ to realize significant modulation of the probe beam during its propagation, and the pump beam should be with a frequency higher than the plasma frequency. Therefore, one can create a PZP designed to manipulate lasers of longer wavelength with a pump beam of a shorter wavelength and larger spotsize with an extremely underdense plasma. For instance, as it is challenging to obtain high-power terahertz (THz) lasers in the laboratory, one can use a pump beam of a wavelength on the order of micrometers and a beam waist on the order of millimeters, to modulate an extremely underdense plasma whose density is set to be on the order of the critical density of THz lasers to produce a zone plate for THz laser manipulation. As an example, a PIC simulation is conducted where a $10\mu m$ LG laser is launched into a homogeneous plasma of density $n_{e0}=3.35\times{10}^{22}m^{-3}$ to create a THz PZP. A THz laser ($\lambda_b = 100\mu m,\nu=3THz$) is launched as the probe beam and the focusing result on the optical axis is shown by the purple dash-dotted line in figure 3(a). Here the difference in wavelength between the pump beam and the probe beam in the simulation is only 10 times due to the limited computing resources. It can be much more significant to meet the demands of practical applications. The given scheme may pave a new way for THz and visible laser manipulation at higher intensities.

In previous simulations, both the pump beam and the probe beam are normally incident onto the plasma slab for demonstration. However, traditional FZP bears a certain robustness to the incident angle, which means that an obliquely incident probe beam within a specific range of incident angle produces a tilted laser field \cite{RN37}.We carried out a contrast simulation launching an obliquely incident probe beam with an incident angle of $15^\circ$ into the PZP obtained in section 2. The intensity distribution in the longitudinal plane is shown in figure 3(b), compared to that of the normally incident case as shown in figure 2(d). The result shows that the foci are tilted with the incident angle while the focal lengths and magnifications remain almost the same as the normally incident case, which is consistent with the traditional FZP. 

\section{Conclusions}
In this manuscript, we have shown that an LG beam with a nonzero $p$ can be used to generate a plasma zone plate (PZP) for high-power laser focusing. The formation of PZP was divided into three stages according to the physical characteristics, and formulae were derived for calculating the density modulation based on the fluid model. We verified our model by PIC and numerical simulations and obtained an annular density distribution composed of alternating rings of high and low densities. The maximum density can be higher than the critical density while the minimum density is close to zero. 

The alternating rings of high and low densities imply alternating rings of low and high refractive index for the probe beam. Therefore, the plasma structure we obtained may be used as a zone plate to focus high-power lasers. We verified the function of PZP by PIC simulations with both normally and obliquely incident probe beams, and obtained multi-focal intensity distributions consistent with the characteristics of the traditional FZP. The intensity is magnified up to more than two orders of magnitude in the simulations. Furthermore, it is demonstrated that PZP is also applicable for focusing terahertz (THz) sources.

The proposed PZP is a new addition to the plasma-based class of optical components for manipulating high-power lasers. The multi-focal laser field achieved owns potential applications to various scenarios, such as laser drilling, microscopy and particle acceleration. 

\ack

This research was supported by the National Natural Science Foundation of China (NSFC) under Grant No. 11975014, by the Strategic Priority Research Program of Chinese Academy of Sciences, Grant Nos. XDA25050400 and XDA25010200. The numerical calculations in this paper have been done on the supercomputing system in the Supercomputing Center of University of Science and Technology of China.

\section*{References}                     

\bibliographystyle{unsrt}

\end{document}